\documentclass[12pt, letterpaper, ]{article}

\usepackage[utf8]{inputenc}
\usepackage{amssymb}
\usepackage{multicol}
\usepackage[margin=1.9cm]{geometry}
\usepackage{setspace}
\usepackage{graphicx}
\usepackage{url}
\usepackage{color}
\usepackage{hyperref}
\graphicspath{ {images/} }
\usepackage{tikz}
\usepackage{amsmath}

\begin{document}
	
	\title{Distribution can be Dropped: Reply to Rumfitt}
	
	\author{Iulian D. Toader\thanks{Institute Vienna Circle, University of Vienna, Vienna, Austria.  iulian.danut.toader@univie.ac.at}}
	
	\date{(forthcoming in \textit{Analysis})}
	
	\maketitle
	
	\begin{abstract}
		
		Most believe that there are no empirical grounds that make the adoption of quantum logic necessary. Ian Rumfitt has further argued that this adoption is not possible, either, for the proof that distribution fails in quantum mechanics is rule-circular or unsound. I respond to Rumfitt, by showing that neither is the case: rule-circularity disappears when an appropriate semantics is considered, and soundness is restored by slightly modifying standard quantum mechanics. Thus, albeit this is indeed not necessary, it is however possible for a quantum logician to rationally adjudicate against classical logic.
		
		\bigskip
		
		\textbf{Keywords}: truth-ground semantics, quantum logic, quantum mechanics, rule-circularity 
		
	\end{abstract}

	\section{\textit{Introduction}}
	
	The transition from classical to quantum mechanics (QM) was taken by Birkhoff and von Neumann (1936) to require a replacement of classical logic (CL) by a non-distributive quantum logic (QL). There is nowadays a consensus that Putnam's arguments (1968) that QL must be accepted on empirical grounds have been unsuccessful. A more recent objection, however, due to Rumfitt (2015), points out that one cannot accept QL on rational grounds, either. For one cannot justify the failure of distribution in QM without assuming this very rule in the metalanguage. This objection can be answered by pointing out that it equivocates on the meaning of ``distribution'', for as Fine rightly emphasized, ``the sense of the distributive law in which it is said to fail is not the sense in which, as the distributive law, it is supposed to hold.'' (Fine 1972: 19; see Horvat and Toader 2024 for discussion). But even if one ignored this equivocation, there is an alternative way to resist Rumfitt's objection, as I will show in this paper.
	
	\section{\textit{The Proof}}
	
	Birkhoff and von Neumann's argument for replacing the package QM \& CL by the alternative package QM \& QL does not explicitly say anything about metaphysics. Taking this up in the late 1960s, Putnam argued for replacing QM \& CL \& IM by QM \& QL \& TM, where intolerable metaphysics is part of the former package and tolerable metaphysics part of the latter. He pointed out that QM \& QL \& TM is preferable to QM \& CL \& IM, since the absence of worldly indeterminacy is more tolerable than its presence, just as the absence of hidden variables is more tolerable than their presence, on any measure of tolerance for realist metaphysical hypotheses. In addition, QM \& CL \& IM makes stronger logical assumptions, since it demands that conjunction distribute over disjunction. On this basis, Putnam defended the view that CL must be revised by QL, holding on to this view for almost 25 years. Later on, Putnam came to acknowledge packages that preserve CL and include realist metaphysical hypotheses that are not that intolerable: ``Surely, before we accept views that require us to revise our logic, we need to be sure that it is \textit{necessary} to go \textit{that} far to make sense of quantum phenomena. And we now know that it is not. ... I was wrong to think that a tenable realistic interpretation must give up classical logic.`` (Putnam 2012: 175-177) 
	
	However, it remains unclear whether Putnam was ready to concede more than this, and in particular that he was perhaps also wrong to think that it is even \textit{possible} to go that far, that a tenable realist interpretation of quantum mechanics \textit{can} give up CL and adopt QL instead. Of course, one might think that the possibility of adopting QL in QM should not hang on the tenability of a realist interpretation, especially if -- as some philosophers claim -- a QL realist interpretation of QM has already been shown untenable and is now of merely historical interest (Maudlin 2010). Indeed, as announced above, my focus in this section will be on Rumfitt's objection, which can be analyzed independently of the issue of the tenability of a realist interpretation, that Putnam was wrong to believe that the rule of distribution can be ditched. To facilitate the analysis of this objection, this section offers a formal proof that distribution fails in QM, as well as its validation by the semantics assumed by Putnam.
	
	Let $\mathcal{QL}$ be the quantum logical calculus, which includes a sentential language, $L$, with variables for atomic sentences, $p$, $q$, ..., and symbols for logical connectives, $\neg, \wedge, \vee $, such that the set $S_\mathcal{QL}$ of sentences is defined inductively. Let $\mathcal{QL}$ have most rules of a classical natural deduction system, such as double negation, De Morgan rules, as well as introduction and elimination rules for all connectives (with the exception of unrestricted $\vee$-elimination).\footnote{Of course, $\mathcal{QL}$ might also include \textit{modus ponens} and \textit{modus tollens}, with a counterfactual conditional, like the so-called Sasaki hook, defined as follows: $ p \rightarrow q $ iff $ \neg p \vee ( p \wedge q ) $. Nothing in the paper hangs on this, however.} For any $i \in \{1, ..., n\}$, let $A_{i}$ be sentences in $S_\mathcal{QL}$ stating the possible positions of a quantum system \textit{P}, and for any $j \in \{1, ..., m\}$, let $B_{j}$ be sentences in $S_\mathcal{QL}$ stating its possible momenta. Assume that the state space associated with \textit{P} is a finite-dimensional Hilbert space, and assume the standard eigenstate-eigenvalue link (EEL), i.e., that \textit{P}'s observables can have precise values only at their eigenstates. Suppose \textit{P} is in a position eigenstate, and $A_{z}$ states \textit{P}'s measured position. The argument that distribution fails can be formalized by the following derivation (henceforth referred to as ``the Proof''):
	
	\bigskip
	
	$ \begin{array}{llr}
		
		1.\  A_{z} \wedge (B_{1} \vee ... \vee B_{m}) & \text{premise} \\
		
		2.\  \neg (A_{i} \wedge B_{j}) & \text{premise} \\
		
		3.\  A_{z} & 1, \wedge\text{-elimination} \\
		
		4.\  \neg(A_{z} \wedge B_{1}) \wedge ... \wedge \neg (A_{z} \wedge B_{m}) & 2, 3, \text{substitution}, \wedge\text{-introduction} \\
		
		5.\  \neg( (A_{z} \wedge B_{1}) \vee ... \vee (A_{z} \wedge B_{m})) & 4, \text{de Morgan}
		
	\end{array} $
	
	\bigskip
	
	The Proof provides us with a counterexample to the rule of distribution in \textit{L}, which can be stated as $ \ulcorner A_{z} \wedge (B_{1} \vee ... \vee B_{m}) \urcorner \nvdash \ulcorner (A_{z} \wedge B_{1}) \vee ... \vee (A_{z} \wedge B_{m})\urcorner $. To see why Putnam took this counterexample to be validated by \textit{L}'s semantics, let's have a look at the semantics he considered.
	
	Let $\mathcal{OO}$ be an orthomodular ortholattice, i.e., a partially ordered set $S_\mathcal{OO}$ of elements, such as the closed subspaces of the Hilbert space associated with \textit{P}, together with operations $^{\bot}, \cap, \cup$, for orthocomplementation, meet, and join (or span), respectively. This lattice satisfies the rule of orthomodularity -- an even weaker form of distribution than modularity -- which states that, for any $a, b \in S_\mathcal{OO}$,
	
	\begin{center}
		
		if $a \subseteq b$, then $b = a \cup (a^{\bot} \cap b)$. 
		
	\end{center}
	
	We say that $\mathcal{OO}$ is a model of $\mathcal{QL}$ if and only if there is a map $h : S_\mathcal{QL} \longrightarrow S_\mathcal{OO}$ such that \textit{h} is a homomorphism:
	
	\begin{center}
		
		$h(\neg p) = h(p)^{\bot}$ \\ 
		$h(p \wedge q) = h(p) \cap h(q)$ \\ 
		$h(p \vee q) = h(p) \cup h(q) $ 
		
	\end{center}
	
	On the basis of this homomorphism and the partial order of $\mathcal{OO}$, one can then define orthocomplementation as set-theoretical complementation plus partial order reversal, and logical consequence:
	
	\begin{center}
		
		$h(p) = h(q)^{\bot}$ iff $\{x : x \subseteq h(p)\} = \{x : x {\bot} h(q) \}$
		
		$\Gamma \models_\mathcal{QL} q$ iff $\cap \{h(p):p \in \Gamma \} \subseteq h(q)$.
	\end{center}
	
	The question is whether $\mathcal{OO}$ verifies the Proof. Does it validate the above counterexample to distribution? Indeed, let $\varnothing$ be the empty space and \textbf{1} be the entire Hilbert space. Then the validation goes as follows:
	
	\begin{center}
		
		$ h(A_{z}) \cap (h(B_{1}) \cup ... \cup h(B_{m})) \nsubseteq (h(A_{z}) \cap h(B_{1})) \cup ... \cup (h(A_{z}) \cap h(B_{m})) $ 
		
		$ h(A_{z}) \cap \textbf{1} \nsubseteq \varnothing \cup ... \cup \varnothing $ 
		
		$ h(A_{z}) \nsubseteq  \varnothing $. 
		
	\end{center}
	
	The failure of distribution is a direct consequence of the fact that the sentences in $S_\mathcal{QL}$, interpreted as experimental propositions in QM, form the non-distributive lattice $\mathcal{OO}$, which can be arguably understood as a consequence of the non-commutativity of the algebra of quantum-mechanical observables. The most important thing to note, for present purposes, is that $\mathcal{OO}$ makes the QL connectives non-truth-functional.\footnote{For an in-depth discussion of the non-truth-functionality of QL connectives, see Horvat and Toader 2024. Non-truth-functionality arguably follows from the Kochen-Specker theorem as well, which guarantees that the partial Boolean algebra of the Hilbert space has no homomorphic Boolean extension, i.e., there is no homomorphism from the partial Boolean algebra to the two-valued Boolean algebra $\{0,1\}$, on the assumption that the space is of dimension $d>2$ (Kochen and Specker 1967). For some discussion, see Dickson 1998, especially section 4.1.2.} But Putnam does not seem to have been worried at all about non-truth-functionality, or in any case did not think that this had any implications with respect to the semantic attributes of QL connectives, for he famously, though controversially, claimed that ``adopting quantum logic is not changing the meaning of the logical connectives, but merely changing our minds about the [distributive] law.'' (Putnam 1968: 190) 
	
	Non-truth-functionality is, however, the very reason why Rumfitt rejects $\mathcal{OO}$ as a semantics without any logical significance:
	
	\begin{quote} 
		
		The operations of intersection, span, and orthocomplement on the subspaces of a Hilbert space indeed form a non-distributive lattice, and that non-distributive lattice is useful in quantum-theoretical calculations. But we have as yet been given no reason to assign any \textit{logical} significance to the lattice. To do so, it would need to be argued that, when the statement \textit{A} is true at precisely the states in \textit{h(A)}, and when the statement \textit{B} is true at precisely the states in \textit{h(B)}, then the disjunction $\ulcorner A \vee B \urcorner $ is true at precisely the states in the span of \textit{h(A)} and \textit{h(B)}. Putnam gives no such argument, so his claim that the proposed rules have any logical significance is unsupported.'' (Rumfitt 2015: 176; modified for uniform notation)
		
	\end{quote}
	
	What would need to be argued, according to Rumfitt, is precisely that QL disjunction is truth-functional. For he seems to want not only that $h : S_\mathcal{QL} \longrightarrow S_\mathcal{OO}$ is a homomorphism, but also that $g : S_\mathcal{OO} \longrightarrow \{0,1\}$ is a homomorphism as well. But this would entail that any truth valuation is a homomorphism $v : S_\mathcal{QL} \longrightarrow \{0,1\}$, since $v = g \circ h$. In the case of QL disjunction, this would entail that 
	
	\begin{center}
		
		$v(A \vee B) =1$ if and only if $v(p) = v(q) = 1$.
		
	\end{center}
	
	Thus, what Rumfitt would have wanted Putnam to argue is that QL disjunction is truth-functional. Since Putnam did not so argue, for it is actually impossible to do so, Rumfitt rejected $\mathcal{OO}$ as a logically insignificant semantics. But this raises a difficult question in the philosophy of logic: what justifies the assignment of logical significance to a certain semantics? More specifically, why should logical significance be dictated, as Rumfitt appears to assume, by our pre-theoretical intuitions about truth-functionality, as opposed to our most successful scientific theories? Leaving this for later, I want now to discuss his objection against the Proof.

	\section{\textit{Removing Rule-Circularity}}
	
	Rumfitt holds the general view that while simplicity and strength are significant, perhaps even predominant criteria for choosing between rival scientific theories, such criteria are not indispensable when it comes to choosing between rival logics. He follows Dummett's view on the conditions for the mutual understanding between logicians of different denominations: ``How can the classical logician and the non-standard logician come to understand one another? Not, obviously, by defining the logical constants. They have to give a semantic theory; and they need one as stable as possible under changes in the underlying logic of the metalanguage.'' (Dummett 1987: 254) Rumfitt takes Dummett to have ``pointed the way'' towards identifying what is necessary for rational adjudication between logics (Rumfitt 2015: 9). Thus, in the case of interest here, if $\mathcal{QL}$ were given a classical semantics, then the Proof cannot adjudicate against CL, for a classical semantics would not be stable enough if one adopted QL (rather than CL) in the metalanguage. If one did adopt CL in the metalanguage, the Proof still cannot adjudicate against CL, because a non-classical semantics for $\mathcal{QL}$ would be also unstable (for reasons we are about to discuss). These violations of Dummett's stability condition justify the classical logician's rejection of the Proof on rational grounds, or so Rumfitt believes. But I don't think that his view is correct, and in this section I want to provide some reasons for doubting it's tenability. 
	
	Let $\mathcal{TG}_{1}$ be a non-classical semantics, what Rumfitt calls a truth-ground semantics. The basic notion of this semantics is that of a truth-ground for a statement, that is a closed set of possibilities at which the statement is true. A set of possibilities, \textit{U}, is closed if and only if it is its own closure, \textit{Cl(U)}, where this is ``the smallest set of possibilities at which every statement that is true throughout \textit{U} is true.'' (Rumfitt 2015: 162) Let $S_{\mathcal{TG}_{1}}$ be a set of truth-grounds, together with operations $^{\bot}, \cap, \cup$, for orthocomplementation, intersection, and union, respectively. Then $\mathcal{TG}_{1}$ is a model of $\mathcal{QL}$ if and only if the truth-grounds can be taken as the closed subspaces of a finite-dimensional Hilbert space associated with our physical system \textit{P} and there is a homomorphism $r_{1} : S_\mathcal{QL} \longrightarrow S_{\mathcal{TG}_{1}}$ such that:
	
	\begin{center}
		
		$r_{1}(\neg p) = r_{1}(p)^{\bot}$ \\ $r_{1}(p \wedge q) = r_{1}(p) \cap r_{1}(q)$ \\ $r_{1}(p \vee q) = Cl(r_{1}(p) \cup r_{1}(q)) $ 
		
	\end{center}
	
	The closure operation, \textit{Cl}, which is taken here as a primitive relation, has some nice lattice-theoretical properties such as
	
	\begin{center}
		
		$ r_{1}(p) \subseteq Cl(r_{1}(p))$, \\ $ClCl(r_{1}(p))=Cl(r_{1}(p))$, and  \\ if $r_{1}(p) \subseteq r_{1}(q)$, then $Cl(r_{1}(p)) \subseteq Cl(r_{1}(q))$     
	\end{center}
	
	Importantly, however, \textit{Cl} does not have the following topological property:
	
	\begin{center}
		
		$Cl(\varnothing \cup ... \cup \varnothing) = Cl(\varnothing)= \varnothing$,     
		
	\end{center}
	
	so as a consequence, it cannot verify that a disjunction of false sentences is false (Rumfitt 2015: 135-6, 162-3). It turns out that, precisely because it lacks this property, $\mathcal{TG}_{1}$ by itself cannot validate our above counterexample to distribution:
	
	\begin{center}
		
		$ r_{1}(A_{z}) \cap Cl((r_{1}(B_{1}) \cup ... \cup r_{1}(B_{m})) \nsubseteq Cl((r_{1}(A_{z}) \cap r_{1}(B_{1})) \cup ... \cup (r_{1}(A_{z}) \cap r_{1}(B_{m}))) $ 
		
		\bigskip
		
		$ r_{1}(A_{z}) \cap Cl(\varnothing \cup ... \cup \varnothing)) \nsubseteq Cl(\varnothing \cup ... \cup \varnothing) $
		
	\end{center}
	
	In order to be able to validate the counterexample, $\mathcal{TG}_{1}$ must be supplemented by a metalogical proof that a disjunction of false sentences is false, i.e., a proof in the metalanguage to the effect that, on line 5 of the Proof, disjunction is truth-functional. As Rumfitt puts it, one needs a proof of ``the semantic principle that a true disjunction must contain at least one true disjunct'' (Rumfitt 2015: 174-5). This is supposed to make up for $\mathcal{TG}_{1}$'s missing the above topological property.
	
	However, Rumfitt notes, the metalogical proof requires the unrestricted rule of $\vee$-elimination and, thus, assumes CL. The Proof is, therefore, rule-circular (since the unrestricted rule of $\vee$-elimination is logically equivalent to distribution). Because of this rule-circularity, Rumfitt rejected $\mathcal{TG}_{1}$ as an unacceptable semantics for $\mathcal{QL}$ on rational grounds. For convenience, here is the metalogical proof (including a tacit correction of Rumfitt's justification for its last step): 
	
	\bigskip
	
	$\begin{array}{lll}
		
		1.\ Tr(\ulcorner A \vee B\urcorner) & \text{premise} \\
		
		2.\ Tr(\ulcorner A \vee B \urcorner) \longrightarrow A \vee B & \text{principle about truth} \\
		
		3.\ A \vee B & \text{1, 2, modus ponens} \\
		
		4.\ A & \text{assumption} \\
		
		5.\ A \longrightarrow Tr(\ulcorner A \urcorner) & \text{principle about truth} & \text{(side premise)} \\
		
		6.\ Tr(\ulcorner A \urcorner) & \text{4, 5, modus ponens} \\
		
		7.\ Tr(\ulcorner A \urcorner) \vee Tr(\ulcorner B \urcorner) & \text{6,} \vee\text{-introduction} \\
		
		8.\ B & \text{assumption} \\
		
		9.\ B \longrightarrow Tr(\ulcorner B \urcorner) & \text{principle about truth} & \text{(side premise)} \\
		
		10.\ Tr(\ulcorner B \urcorner) & \text{8, 9, modus ponens} \\
		
		11.\ Tr(\ulcorner A \urcorner) \vee Tr(\ulcorner B \urcorner) & \text{10,} \vee\text{-introduction} \\
		
		12.\ Tr(\ulcorner A \urcorner) \vee Tr(\ulcorner B \urcorner) & \text{3, 5, 9,} \vee\text{-elimination}
		
	\end{array}$
	
	\bigskip
	
	Note, first, that since the calculus $\mathcal{QL}$ is not, itself, a quantum system, there can be no surprise that the metalanguage has a truth-functional and, thus, classical  disjunction. Assuming that \textit{L} is a countable language, one could also prove that the semantics of the metalanguage of \textit{L} is a distributive lattice, based on Dunn's classical recapture of distribution as an arithmetical theorem in $PA_{QL}^{1}$, i.e., first-order Peano arithmetic with QL (Dunn 1980). Of course, one would expect the classical logician to insist that, despite such classical recapture results, one should still not use CL in an argument against CL. Secondly, recall that we only needed a classical metalogical proof because $\mathcal{TG}_{1}$ was unable, by itself, to validate Putnam's proof. But a quantum logician can, and I think should, reject this semantics. 
	
	Rumffit assumes that $\mathcal{TG}_{1}$ is ``acceptable to adherents of many rival logical schools; these principles [i.e., the three semantic rules for logical connectives], then, have a good claim to articulate the commonly understood senses of the sentential connectives.'' (Rumfitt 2015: 167) Good, but not good enough. In order to have a better claim to articulate the semantic attributes of QL connectives, and in particular those of QL disjunction, $\mathcal{TG}_{1}$ must be revised. In particular, instead of being taken as a primitive, the \textit{Cl} operation on the set of truth-grounds must be defined in terms of quantum incompatibility, i.e., the relation that obtains between quantum observables in virtue of the fact that the operators representing them on the Hilbert space associated with \textit{P} do not commute. I turn now to considering Rumfitt's second, revised truth-ground semantics for $\mathcal{QL}$.
	
	Let $\mathcal{TG}_{2}$ be the revised truth-ground semantics, with a set $S_{\mathcal{TG}_2}$ of truth-grounds and operations on them, $^{\bot}, \cap, \cup$, for orthocomplementation, intersection, and union, respectively. Then $\mathcal{TG}_{2}$ is a model of $\mathcal{QL}$ if and only if the truth-grounds are the closed subspaces of a finite-dimensional Hilbert space associated with system \textit{P} and there is a map $r_{2} : S_\mathcal{QL} \longrightarrow S_{\mathcal{TG}_2}$ such that $r_{2}$ is a homomorphism:
	
	\begin{center}
		
		$r_{2}(\neg p) = r_{2}(p)^{\bot}$ \\ $r_{2}(p \wedge q) = r_{2}(p) \cap r_{2}(q)$ \\ $r_{2}(p \vee q) = (r_{2}(p) \cup r_{2}(q))^{\bot \bot} $ 
		
	\end{center}
	
	Orthocomplementation and logical consequence are, as before, defined lattice-theoretically in terms of partial order. But the \textit{Cl} operation is now defined as double orthocomplementation: $Cl(r_{2}(p)) = r_{2}(p)^{\bot \bot}$. Furthermore, orthocomplementation has the topological property that was lacking in the previous $\mathcal{TG}_{1}$ semantics:
	
	\begin{center}
		
		$(\varnothing \cup ... \cup \varnothing)^{\bot \bot} = \varnothing^{\bot \bot} = \varnothing$
		
	\end{center}
	
	It would seem that precisely because orthocomplementation has this property, the truth-functionality of disjunction on line 5 of the Proof can be justified, so then $\mathcal{TG}_{2}$ could validate our counterexample to distribution:
	
	\begin{center}
		
		$ r_{2}(A_{z}) \cap (r_{2}(B_{1}) \cup ... \cup r_{2}(B_{m}))^{\bot \bot} \nsubseteq ((r_{2}(A_{z}) \cap r_{2}(B_{1})) \cup ... \cup (r_{2}(A_{z}) \cap r_{2}(B_{m})))^{\bot \bot} $ 
		
		$ r_{2}(A_{z}) \cap (\varnothing \cup ... \cup \varnothing)^{\bot \bot} \nsubseteq  (\varnothing \cup ... \cup \varnothing)^{\bot \bot} $
		
		$ r_{2}(A_{z}) \nsubseteq \varnothing $
		
	\end{center}
	
	Yet, Rumfitt argues that this is still not the case, despite the fact that no metalogical proof of the truth-functionality of disjunction is needed any longer, so that rule-circularity is now avoided. The reason given for $\mathcal{TG}_{2}$'s inability to validate the counterexample to distribution is that, on the first line of the Proof, $A_{z}$ is false, since $r_{2}(A_{z}) = \varnothing$. 
	
	In order to see why Rumfitt maintains that $r_{2}(A_{z}) = \varnothing$, a justified revision of the Proof is needed. What he believes justifies the revision are some features of standard QM that Putnam professed to ignore (Putnam 1968: 178). More exactly, Rumfitt notes the following: 
	
	\begin{quote} 
		
		The truth principle [i.e., the eigenstate-eigenvalue link] tells us that a statement ... that attributes a precise value to an observable quantity, will be true only at eigenstates of that observable. In a finite-dimensional Hilbert space, any self-adjoint operator will possess eigenstates, so in such a space there will be a state at which [such a statement] is true. In an infinite-dimensional space, however, even self-adjoint operators need have no eigenstates at all. ... once an infinite-dimensional Hilbert space is endowed with observables for position and momentum, there are no eigenvalues for position. (Rumfitt 2015: 180)
		
	\end{quote}
	
	Indeed, on an infinite-dimensional Hilbert space, those operators corresponding to observables with continuous spectra, like position and momentum, have no eigenstates in a Schrödinger representation, and thus such observables can have no precise values. As a consequence, the Proof has to be revised. Needless to say, the revision does not make the (unrevised) Proof superfluous. Since QM makes extensive use of finite-dimensional Hilbert spaces in modeling observables like spin, which have discrete spectra, the Proof is enough to provide a counterexample to distribution. The revision is only needed if one looks for a counterexample that involves observables with continuous spectra.

	\section{\textit{Restoring Soundness}}
	
	Let $\mathcal{QL}$ be our calculus, as before. Assume, again, that the EEL is true, but let the state space associated with our quantum system \textit{P} be an infinite-dimensional Hilbert space. For any $i \in \{1, ..., n, ...\}$, let $A_{i} \in S_{\mathcal{QL}}$ be sentences that state the possible positions of \textit{P}, and for any $j \in \{1, ..., m, ...\}$, let $B_{j} \in S_{\mathcal{QL}}$ be sentences that state its possible momenta. Then the revised Proof is the following derivation:
	
	\bigskip
	
	$\begin{array}{ll}
		
		1.\ A_{z} \wedge (B_{1} \vee ... \vee B_{m} \vee ...) & \text{premise} \\
		
		2.\ \neg (A_{i} \wedge B_{j}) & \text{premise} \\
		
		3.\ A_{z} & \text{1,} \wedge\text{-elimination} \\
		
		4.\ \neg(A_{z} \wedge B_{1}) \wedge ... \wedge \neg (A_{z} \wedge B_{m}) \wedge ... & \text{2, 3, substitution,} \wedge\text{-introduction} \\
		
		5.\ \neg( (A_{z} \wedge B_{1}) \vee ... \vee (A_{z} \wedge B_{m}) \vee ...) &  \text{4, de Morgan}
		
	\end{array}$
	
	\bigskip
	
	This derivation provides us with a counterexample to distribution in \textit{L}, which can be stated as $\ulcorner A_{z} \wedge (B_{1} \vee ... \vee B_{m} \vee ...) \urcorner \nvdash \ulcorner (A_{z} \wedge B_{1}) \vee ... \vee (A_{z} \wedge B_{m}) \vee ... \urcorner$. This counterexample is, however, not validated by $\mathcal{TG}_{2}$:
	
	\begin{center}
		
		$ r_{2}(A_{z}) \cap (r_{2}(B_{1}) \cup ... \cup r_{2}(B_{m}) \cup ...)^{\bot \bot} \nsubseteq ((r_{2}(A_{z}) \cap r_{2}(B_{1})) \cup ... \cup (r_{2}(A_{z}) \cap r_{2}(B_{m})) \cup ...)^{\bot \bot} $ 
		
		$ r_{2}(A_{z}) \cap \textbf{1} \nsubseteq (\varnothing \cup ... \cup \varnothing \cup ...)^{\bot \bot} $
		
		$ \varnothing \nsubseteq \varnothing $
		
	\end{center}
	
	As already pointed out, $ r_{2}(A_{z}) = \varnothing $ because \textit{P}'s position has no eigenstates. This entails that $A_{z}$ is false and, thus, that the revised Proof is unsound with respect to $\mathcal{TG}_{2}$. 
	
	However, the soundness of the revised Proof might easily be restored by simply denying the EEL. This is not unheard of, and some have contended that denying the EEL is fully justified, since it ``has nothing much to do with quantum theory. ... It’s an interpretive assumption. Its motivation (I think) comes from the idea that measurement must be discovering some preexisting measured value, in which case that value must be possessed by a system iff it is certain to give that value as a result of measurement.  But this isn’t realized in any realist interpretation of quantum mechanics ... And it is anyway incompatible with the actual physics of quantities with continuously many measurement outcome possibilities, like position and momentum.'' (Wallace 2013: 215) Of course, this view conflicts with the Kochen-Specker theorem, and the conflict resolves by dropping distribution (Demopoulos 1976). But this solution is not available here, since whether distribution can be dropped is the very problem that denying the EEL attempts to address. 
	
	Less radically perhaps, one could also just revise the EEL, so that instead of allowing that observables have precise values at all states, one modifies the notion of a state such that observables like position and momentum have precise values even though the state space of the system is an infinite-dimensional Hilbert space. This could be implemented, for example, by coarse-graining the state space, an operation which allows that, for system \textit{P}, an observable $O$ with a continuous spectrum can have precise values when the state of \textit{P} assigns probability 1 to some cells or regions, rather than to points, in that space. For any such observable, these regions are determined, say, by the interval $[x - \epsilon, x + \epsilon]$, for some $\epsilon > 0$ and for any possible value \textit{x} in the spectrum of that observable. The revised EEL then stipulates that a physical system $P$ has a precise value for observable $O$ if and only if the state of $P$ is \textit{a coarse-grained state} of $O$ (Fine 1971, Teller 1979, Halvorson 2001). Consequently, the statement $A_{z}$ on line 1 of the revised Proof will be true at some coarse-grained position state. Thus, the revised Proof is not unsound after all. 
	
	Rumfitt considers a similar revision of the EEL, but rejects it as an ``unsatisfactory constraint on truth'' (Rumfitt 2015: 180). \textit{Mutatis mutandis}, the revision of the EEL that I just suggested would also be considered unsatisfactory, on the ground that having a precise position at a coarse-grained state really amounts to nothing else than having an imprecise position. For if the statement $A_{z}$ is taken to describe a precise position of \textit{P}, then it can be only approximately true. But since logic is concerned with the preservation of truth, not that of approximate truth, revising the EEL in the way coarse-graining allows would change the subject of logic. However, in response to this, one can surely revise the EEL even further, so that it stipulates that an observable $O$ of $P$ has \textit{a coarse-grained value} if and only if the state of $P$ is a coarse-grained state of $O$. In this case then, when $A_{z}$ is taken to describe a coarse-grained position of \textit{P}, the statement $A_{z}$ will be true, rather than approximately true, at some coarse-grained position states. Therefore, coarse-graining the infinite-dimensional Hilbert space and then revising the EEL accordingly can successfully block Rumfitt's argument against the revised Proof. This is done \textit{without changing the subject of logic}, just as he correctly demands. 
	
	My response admittedly requires some slight changes to the standard formalism of QM, e.g., an application of mathematical notions from coarse geometry on metric spaces (Roe 2003). But one might object that the response is beside the point, since the (revised) Proof assumes standard QM. However, my goal is not to defend standard QM. Rather, it is to show that rational adjudication against CL in QM is possible. If this requires moving beyond standard QM, that's fine by me. After all, classical logicians, too, usually go beyond standard QM, invoking one of its dynamical extensions or interpretations, when they insist that distribution need not be dropped, as indeed Putnam himself ultimately did.
	
	\section{\textit{Conclusion}} 
	
	As noted above, Rumfitt maintains that the orthomodular ortholattice $\mathcal{OO}$ should be rejected as logically insignificant because it makes disjunction not-truth-functional, which raises a fundamental issue: What is a proper justification of our attribution of logical significance to a given semantics? Should such a justification turn to our intuitions and dispositions or to our theories and theorems? Should our considered reasons for assigning logical significance then necessarily align with truth-functionality, or not? My take on this comes close to the view on logic suggested in the following passage: 
	
	\begin{quote} 
		
		What counts as an example of good reasoning, and how do we come to know what those examples are? Do we `just know' one when we see it? No. We must have some way to assign intersubjectively available truth values to sentences involving logical connectives. In other words, we need some way to determine whether a given form of logical inference is `successful'. Given the principle that the world (or, a correct theoretical description of the world) does not `disobey' the correct rules of reasoning, we may hope to determine which inferences are correct by determining
		how to relate propositions involving logical connectives to empirical facts. (Dickson 2001: S283)
		
	\end{quote}
	
	If this view were adopted, then one should not reject $\mathcal{OO}$ as logically insignificant, but rather take the non-truth-functionality of QL connectives seriously. But note that this need not entail a commitment to the claim that QL is universally true, which would in turn entail that the rule of distribution \textit{must} be dropped. Dickson is, in fact, committed to this claim:  
	
	\begin{quote}
		
		Quantum logic is the `true' logic. It plays the role traditionally played by logic, the normative role of determining right-reasoning. Hence the distributive law is wrong. It is not wrong `for quantum systems' or `in the context of physical theories' or anything of the sort. It is just wrong, in the same way that `(p or q) implies p' is wrong. (\textit{op. cit.}, S275)
		
	\end{quote}
	
	However, this is not the view that I have been arguing for. In this paper, I have been interested only in making the point, against Rumfitt, that distribution \textit{can} be dropped, or in other words, that rational adjudication against CL, even though not necessary, is nonetheless possible.\footnote{Rumfitt's problem appears to be a particular instance of what came to be called the adoption problem (Kripke 1974). But I think that if one looks more closely, one can see that it is not. Although this is a longer discussion, which belongs to another paper, here is very briefly how I see this matter. Take the following formulation of the adoption problem: ``certain basic logical principles cannot be \textit{adopted} because, if a subject already infers in accordance with them, no \textit{adoption} is needed, and if the subject does \textit{not} infer in accordance with them, no \textit{adoption} is possible.” (Birman 2024: 39) Thus, QL cannot be adopted because, if a subject already infers non-distributively, no adoption of QL is needed, and if the subject does not infer non-distributively, no adoption of QL is possible. In particular, if the subject infers classically, no adoption of QL is possible. However, Rumfitt's reasons for the impossibility of adopting QL are not simply based on commitment to CL: classicality is \textit{not}, for him, a sufficient condition for rejecting the possibility of adopting QL. In fact, as we have seen above, Rumfitt adjusts the commitment to classicality, as he proposes two non-classical semantics for the evaluation of the Proof. In contrast to Kripke, Rumfitt’s reasons for the impossibility of adopting QL are based on commitment to Dummett’s stability condition: for Rumfitt, stability \textit{is} a necessary condition for the possibility of adopting QL. I thank one referee for pressing this point, and Constantin C. Brincus and Sebastian Horvat for discussion on several other points in this paper.}

	\section*{\textit{References}}
	
	\noindent Birkhoff, G. and J. von Neumann. 1936. The Logic of Quantum Mechanics. \textit{Annals of Mathematics} 
	
	37: 823--843.
	
	\bigskip
	
	\noindent Birman, R. 2024. The Adoption Problem and the Epistemology of Logic. \textit{Mind} 133: 37--60.

	\bigskip
	
	\noindent Demopoulos, W. 1976. The Possibility Structures of Physical Systems. In \textit{Foundations of} 
	
	\textit{Probability Theory, Statistical Inference, and Statistical Theories of Science}, eds. W. L. Harper 
	
	and C. A. Hooker, 55--80, Dordrecht: Reidel.
	
	\bigskip
	
	\noindent Dickson, M. 1998. \textit{Quantum chance and non-locality}. Cambridge: Cambridge University Press
	
	\bigskip
	
	\noindent Dickson, M. 2001. Quantum Logic Is Alive $\wedge$ (It Is True $\vee $ It Is False). \textit{Philosophy of Science} 68: 
	
	S274--S287.
	
	\bigskip
	
	\noindent Dummett, M. 1987. Reply to John McDowell. In \textit{Michael Dummett: Contributions to Philosophy}, 
	
	ed. B. M. Taylor, 253--268, Nijhoff International Philosophy Series, vol. 25, Springer.
	
	\bigskip
	
	\noindent Dunn, J. M. 1980. Quantum Mathematics. \textit{Proceedings of the the Philosophy of Science Association}, 
	
	512--531.
	
	\bigskip
	
	\noindent Fine, A. 1971. Probability in quantum mechanics and in other statistical theories. In \textit{Problems in} 
	
	\textit{the Foundations of Physics}, ed. M. Bunge, 79--92, Springer.
	
	\bigskip
	
	\noindent Fine, A. 1972. Some Conceptual Problems with Quantum Theory. In \textit{Paradigms and Paradoxes:}
	
	\textit{The Philosophical Challenge of the Quantum Domain}, ed. Robert G. Colodny, 3--31, University 
	
	of Pittsburgh Press.
	
	\bigskip
	
	\noindent Halvorson, H. 2001. On the Nature of Continuous Physical Quantities in Classical and Quantum 
	
	Mechanics. \textit{Journal of Philosophical Logic} 30: 27--50.
	
	\bigskip
	
	\noindent Horvat, S. and Toader, I. D. 2023. Quantum Logic and Meaning. arXiv preprint, arXiv:2304.08450.
	
	\bigskip
	
	\noindent Kochen, S. and E. P. Specker. 1967. The Problem of Hidden Variables in Quantum Mechanics. 
	
	In \textit{The Logico-Algebraic Approach to Quantum Mechanics}, I, ed. C. A. Hooker, 293--328, 1975, 
	
	Dordrecht: Reidel.
	
	\bigskip
	
	\noindent Kripke, S. 1974. The Question of Logic. \textit{Mind} 133 (2024): 1--36.
	
	\bigskip
	
	\noindent Maudlin, T. 2010. The Tale of Quantum Logic. In \textit{Hilary Putnam}, ed. Y. Ben-Menahem, 156--187, 
	
	Cambridge: Cambridge University Press
	
	\bigskip
	
	\noindent Putnam, H. 1968. Is Logic Empirical? In \textit{Boston Studies in the Philosophy of Science}, 5, eds. R. S. 
	
	Cohen and M. W. Wartofsky, 216--241, Dordrecht: Reidel. Reprinted as The logic of quantum 
	
	mechanics, in his \textit{Mathematics, Matter and Method. Philosophical Papers}, vol. 1, 174--197, 
	
	Cambridge University Press.
	
	\bigskip
	
	\noindent Putnam, H. 2012. The Curious Story of Quantum Logic. In \textit{Philosophy in the Age of Science:} 
	
	\textit{Physics, Mathematics, and Skepticism}, eds. M. De Caro and D. Macarthur, 162--177, Harvard 
	
	University Press.
	
	\bigskip
	
	\noindent Roe, J. 2003. \textit{Lectures on coarse geometry}. University Lecture Series, American Mathematical 
	
	Society, vol. 31. 
	
	\bigskip
	
	\noindent Rumfitt, I. 2015. \textit{The Boundary Stones of Thought}, Oxford: Oxford University Press.
	
	\bigskip
	
	\noindent Teller, P. 1979. Quantum Mechanics and the Nature of Continuous Physical Quantities. \textit{Journal}
	
	\textit{of Philosophy} 76: 345--361.
	
	\bigskip
	
	\noindent Wallace, D. 2013. A prolegomenon to the ontology of the Everett interpretation. In \textit{The Wave} 
	
	\textit{Function: Essays on the Metaphysics of Quantum Mechanics}, eds. A. Ney and D. Z. Albert, 
	
	203--222, Oxford University Press.
	
\end{document}